\documentclass[preprint,pre,preprintnumbers,superscriptaddress]{revtex4}
\usepackage{latexsym}
\usepackage{amsmath}
\usepackage{graphicx}
\usepackage{bm}
\usepackage{epsf}
\usepackage{epsfig}
\usepackage{dcolumn}

\begin{document}

\title{Strain Engineering Water Transport in Graphene Nano-channels}
\author{Wei Xiong}
\affiliation{Department of Engineering Mechanics and Center for Nano and Micro Mechanics,
Tsinghua University, Beijing 100084, China}
\author{Jefferson Zhe Liu}
\affiliation{Department of Mechanical and Aerospace Engineering, Monash University,
Clayton, VIC 3800, Australia}
\email{zhe.liu@monash.edu}
\author{Ming Ma}
\affiliation{Department of Engineering Mechanics and Center for Nano and Micro Mechanics,
Tsinghua University, Beijing 100084, China}
\author{Zhiping Xu}
\affiliation{Department of Engineering Mechanics and Center for Nano and Micro Mechanics,
Tsinghua University, Beijing 100084, China}
\author{John Sheridan}
\affiliation{Department of Mechanical and Aerospace Engineering, Monash University,
Clayton, VIC 3800, Australia}
\author{Quanshui Zheng}
\affiliation{Department of Engineering Mechanics and Center for Nano and Micro Mechanics, Tsinghua University, Beijing 100084, China}
\affiliation{Institute of Advanced Study, Nanchang University, Nanchang, China}
\email{zhengqs@tsinghua.edu.cn}

\begin{abstract}
Using equilibrium and non-equilibrium molecular dynamic (MD) simulations, we found that engineering the strain on the graphene planes forming a channel can drastically change the interfacial friction of water transport through it. There is a sixfold change of interfacial friction stress when the strain changes from $-10\%$ to $10\%$. Stretching the graphene walls increases the interfacial shear stress, while compressing the graphene walls reduces it. Detailed analysis of the molecular structure reveals the essential roles of the interfacial potential energy barrier and the structural commensurateness between the solid walls and the first water layer. Our results suggest that the strain engineering is an effective way of controlling the water transport inside nano-channels. The resulting quantitative relations between shear stress and slip velocity and the understanding of the molecular mechanisms will be invaluable in designing graphene nano-channel devices.
\end{abstract}

\pacs{Valid PACS appear here}
\date{\today}
\maketitle


\section{Introduction}
Graphene, due to its extremely large specific surface area \cite{1}, superior electronic and mechanical properties, is a promising material in the novel nano-fluidic devices applications, such as, water desalination, nano-filtration, photo-catalysis, super-capacitor etc. \cite{1,2, 3,4,5}. Graphene layers usually self-assemble in paper-like structures with interlayer distance on the nanometer scale \cite{6, 7}. Fluid transport in such flat nano-channels is a key concern in improving the performance of the nano-fluidic devices. Water confined inside nano-channels exhibits superior transport properties owing to its having a different molecular structure than the bulk water \cite{8, 23}. Experiments have shown that the flow rate of water inside carbon nanotubes (CNT) is several orders of magnitude higher than predicted by conventional hydrodynamics \cite{11, 9, 10}. Because of their structural similarity, the extremely fast water transport found inside CNT is also expected in graphene nano-channels. Deep understanding of the molecular transport and the means of controlling the water transport inside the graphene flat nano-channels are highly desired in practice.

The structural properties of the single atomic layer of graphene, to a large extent, are determined by its interactions with the environment. For example, an epitaxial strain of $\sim \pm1\%$ builds up in the graphene when it is grown on different substrates \cite{12, 13, 35, 36}; electromechanical strain on the order of $\pm1-2\%$ results from charge injection in the graphene \cite{14}; and applying mechanical force (e.g., nano-indentation) on the graphene can lead to a strain of about $10\%$ \cite{16}.
%, and applying external electrical field could lead to a strain on the order of $10\%$ in single walled CNT \cite{15}. 
Meanwhile, due to the size confinement in the CNT and graphene nano-channels, the first water layer next to the solid walls tends to be highly ordered \cite{17}. It is reasonable to expect that the slip flow of water over the graphene layers will be analogous to the friction between two ordered crystal planes, meaning the interfacial commensurateness will play an essential role. Strain engineering of graphene layers to alter the interfacial molecular structures could thus provide a method of controlling the flow in nano-fluidic devices.

Here, we use both equilibrium and non-equilibrium MD simulations to study the water flowing in uniformly biaxial strained graphene nano-channels. The relations between the interfacial friction shear stress $\tau$ and the slip velocity $v_s$ are obtained. Our results show that the interfacial friction coefficient (e.g., ratio of $\tau$ over $v_s$) increases almost 6 fold when the strains applied to graphenes varies from $-10\%$ to $10\%$. To gain a deep understanding of the physics behind this remarkable change, the molecular mechanisms, such as the interfacial potential barrier, density and structure factor of the first water layer, are quantitatively investigated.

\section{Simulation details}
\begin{figure}[htbp]
\centerline{\includegraphics[width=0.8\textwidth]{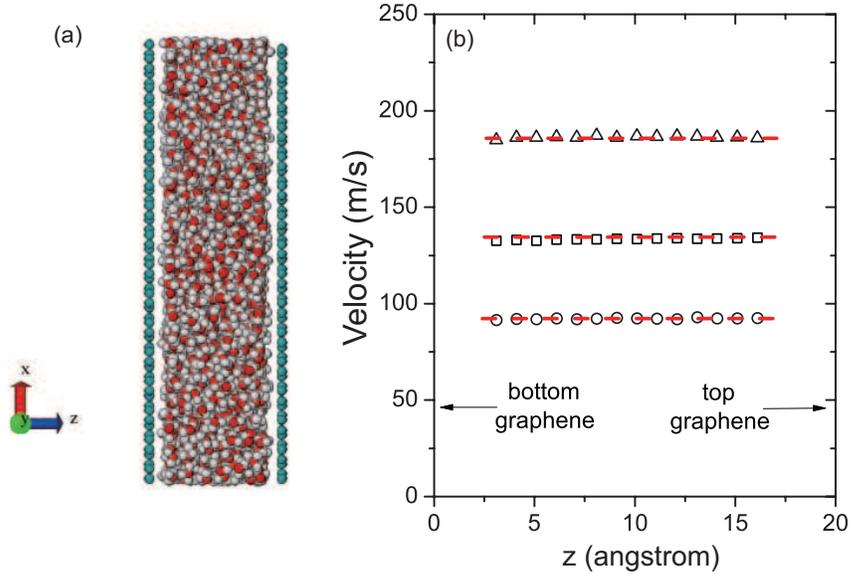}}
\caption{(color online) (a) Water flow inside graphene flat nano-channel with height about 2 nm. 
(b) The Plug-like velocity profile of the Poiseuille flow along the flow direction (i.e., $x$ direction) 
observed in our MD simulations. We applied three different accelerations to water molecules to 
obtain the different steady flow velocities as shown by the different symbols. }
\label{fig1}
\end{figure}

Both equilibrium and non-equilibrium MD simulations were performed. Our molecular system is illustrated in Fig. \ref{fig1}(a): water flows between two graphene sheets with an interlayer distance about 2 nm. Uniform in-plane biaxial strain is applied ranging from $-10\%$ up to $10\%$ to the graphene walls. Accordingly the carbon-carbon bond length $a_{\text{CC}}$ changes from $0.9a_{\text{CC}_0}$ to $1.1a_{\text{CC}_0}$, where $a_{\text{CC}_0}$ denotes the carbon-carbon bond length of strain-free graphene (e.g., $a_{\text{CC}_0} = 0.142$nm). To ensure the internal pressure of water at 1atm, we fix one of the graphene walls and use the other one as a piston to impose the pressure at a given temperature (300K). The position of graphene walls is then fixed in the following MD simulations. Periodic boundary conditions are applied along the two directions in the plane. Along the flow direction ($x$ direction in Fig. \ref{fig1}(a)), the simulation box length was chosen to be 6.0 nm. However, we also performed simulations with a longer length 20 nm for the zero strain case and found the differences in the calculated friction stress were negligible. There are about 1200 to 2000 water molecules in our systems.

All MD simulations were performed with the LAMMPS code \cite{18}. A time step of 1.0 fs was used and the total simulations time was about a few nanoseconds. We used the CHARMM force field and the SPC/E model \cite{19} for water with the SHAKE algorithm \cite{20}. The water-carbon interactions were described by a Lennard-Jones potential between oxygen and carbon atoms with parameters $\epsilon = 4.063$ meV and $\sigma = 0.3190$ nm, yielding a contact angle of $95^o$ between the water and graphene \cite{21}. The van der Waals forces were truncated at 1.2 nm with long-range Columbic interactions computed using the particle-particle particle-mesh (PPPM) algorithm \cite{22}. Water molecules were kept at a constant temperature of 300 K using the Berendsen thermostat, with the temperature calculated after removing the center-of-mass velocity. We also tried the Nose-Hoover thermostat applied to the degrees of freedom perpendicular to the flow direction in the zero-strain case. No significant differences between these two thermostats were found in our study.

In our non-equilibrium MD simulations (NEMD), the Poiseuille flow was driven by applying a constant gravity-like acceleration to all the oxygen and hydrogen atoms. Different external accelerations were applied to achieve a set of steady flows of different velocity (once the external forces were balanced by the friction). It usually took a few hundred picoseconds to reach a steady flow and the simulations were then continued for five more nanoseconds to collect data. By applying different accelerations from 0 to 0.004 nm/ps$^2$, we obtained the steady-flow velocities from 0 to 200 m/s. Velocity profiles of the Poiseuille flows in our nano-channels are plug-like as shown in Fig. \ref{fig1}(b), so we can set the slip velocity equal to the average velocity. We noted that the similar simplification was adopted in Ref \cite{23}. The friction shear stress was calculated from the external force as $\tau = Nma/2A$, where $N$ is the number of water molecules, $m$ is the mass of one water molecule, $a$ is the acceleration, and $A$ is the area of one graphene sheet. For comparison, we also calculated the shear stress $\tau$ using the forces in the flow direction acting on carbon atoms (due to pair-wise interactions between the carbon atoms and the water molecules) and found good agreement, e.g., the difference is less than $5\%$.

\section{Results and Discussions}
\subsection{Strain dependent friction coefficient from MD simulations}

\begin{figure}[htbp]
\centerline{\includegraphics[width=0.6\textwidth]{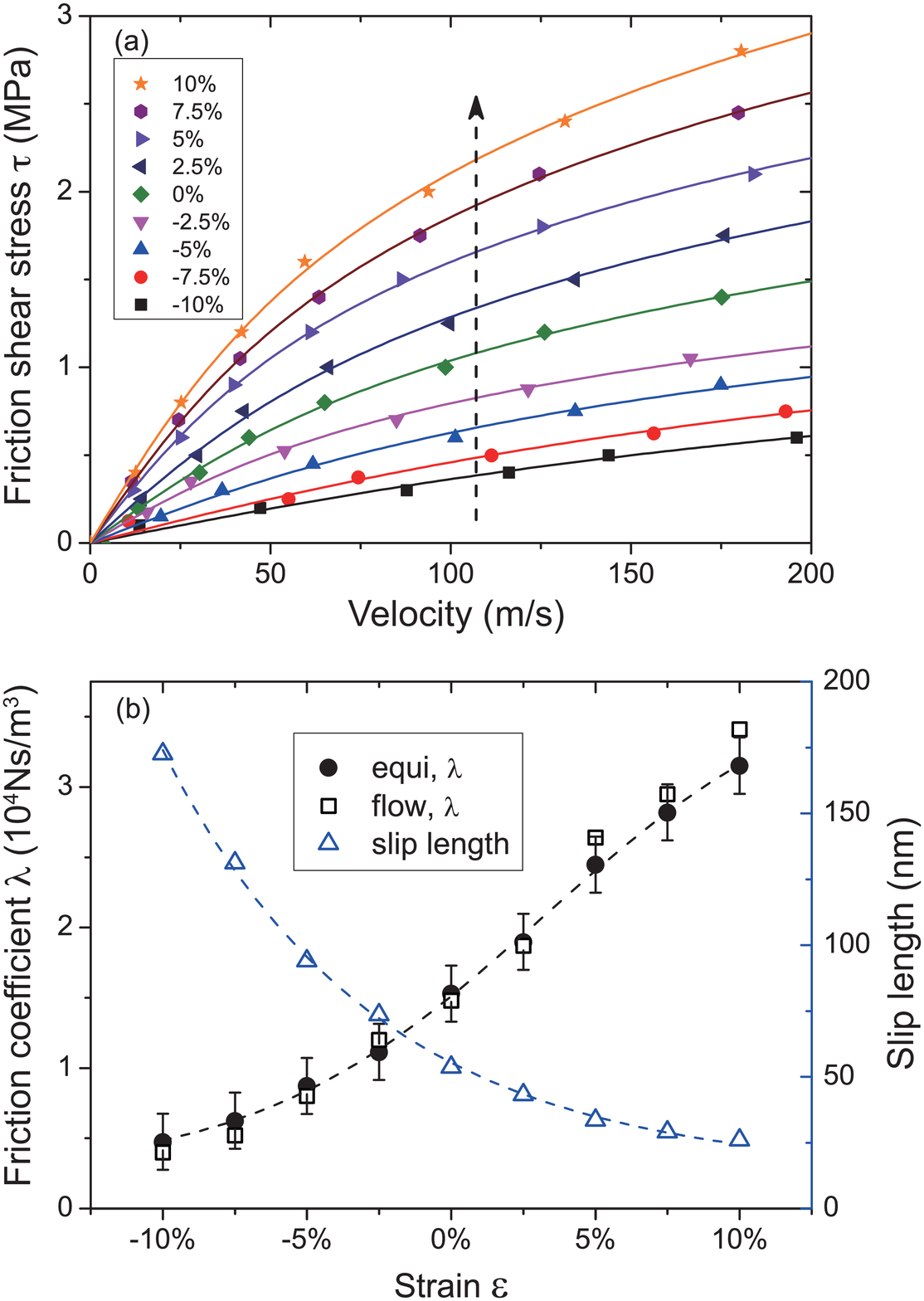}}
\caption{(color online) (a) Friction shear stress $\tau$ versus slip velocity $v_s$ at the water/graphene interface 
with graphene deformation strain $\epsilon$ from $-10\%$ to $10\%$ (from bottom to top). The solid lines represents the fitted relations $\tau/\tau_0 = \text{asinh}(v_s/v_0)$, with the fitting coefficients $\tau_0$ and $v_0$ summarized in Table \ref{table1}. 
(b) The friction coefficient $\lambda$ and the slip length $l_s$ as a function of the applied strain $\epsilon$. 
The open square symbols represent the $\lambda$ values calculated from the slopes of curves at $v_s=0.0$ in (a), 
and the solid circles are calculated by using the GK relation (Eq. (\ref{eq2})) in our equilibrium MD simulations. The slip length $l_s$ is calculated by using $l_s = \mu/\lambda$ \cite{25}, where $\mu$ is the viscosity of water.}
\label{fig2}
\end{figure}

Figure \ref{fig2}(a) shows the interfacial friction shear stress $\tau$ as a function of flow rate $v_s$ in the graphene channels placed under different strain. They are fitted well using an inverse hyperbolic sine (IHS) relationship $\tau/\tau_0 = \text{asinh}(v_s/v_0)$. The values of $\tau_0$ and $v_0$ derived from the fits are summarized in table \ref{table1} for different strain values. The IHS relation's derivation arises from the transition state theory model of Yang \cite{24}, in which the slip flow is described as a collective thermal diffusion of fluid atoms over a periodic solid wall. We have previously justified the use of this relation by using large-scale NEMD simulations for water flowing inside CNTs \cite{23}. Figure \ref{fig2}(a) shows that the friction shear stress increases with increasing of strain.

% Requires the booktabs if the memoir class is not being used
\begin{table*}[htbp]
   \centering
   %\topcaption{Table captions are better up top} % requires the topcapt package
   \caption{Fitted parameters $\tau_0$ and $v_0$ in the inverse hyperbolic-sine relationship, $\tau/\tau_0 = \text{asinh}(v_s /v_0)$, which describes the water slip flow in the strained graphene nano-channels. The friction coefficient $\lambda$ and slip length $l_s$ are also calculated (see text for details).}
   \begin{tabular*}{1.00\textwidth}{@{\extracolsep{\fill}}cccccccccc} % Column formatting, @{} suppresses leading/trailing space
      \toprule
%      \multicolumn{10}{c}{Item} \\
%      \cmidrule(r){1-2} % Partial rule. (r) trims the line a little bit on the right; (l) & (lr) also possible
      $\epsilon$    & $-10\%$ & $-7.5\%$ & $-5.0\%$ & $-2.5\%$ & $0\%$ & $2.5\%$ & $5.0\%$ & $7.5\%$ & $10\%$ \\
      \hline
      $\tau_0$ (MPa)   & 0.457 & 0.528 & 0.507 & 0.487 & 0.688 & 0.827 & 0.880 & 1.059 & 1.183 \\
       $v_0$ (m/s) & 113.6 & 101.5 & 63.4 & 40.7 & 46.5 & 44.2& 33.4 & 35.8 & 34.7 \\
      $\lambda = \tau_0/v_0$  (10$^4$Ns/m$^3$) & 0.40 & 0.52 & 0.80 & 1.20 & 1.48 & 1.87 & 2.64 & 2.96 & 3.41 \\
      $l_s = \mu/\lambda $ (nm)  & 173 & 131 & 94 & 74 & 54 & 43 & 34 & 29 & 26 \\
       \hline\hline
   \end{tabular*}
   \label{table1}
\end{table*}

At low flow rates (e.g., $v_s < 10$m/s), the IHS relationship exhibits a linear relation $\tau \cong \lambda v_s$, where the friction coefficient $\lambda$ is often used to represent the strength of friction. Intrinsically, the friction coefficient is the physically relevant property to characterize the interfacial dynamics \cite{25}. The open squares in Fig. \ref{fig2}(b) depict the friction coefficients $\lambda$ (slopes of the $\tau$ - $v_s$ curves in Fig. \ref{fig2}(a)) as a function of the applied strain $\epsilon$. The key result of Fig. \ref{fig2} is the dramatic effect of the strain on the friction shear stress and the friction coefficient. Stretching the graphene layer leads to the increase of the friction stress with a two fold increase of the friction coefficient at the $10\%$ strain. Compressing the graphene layer results in a significant reduction in the friction coefficient, e.g., about three fold at the $-10\%$ strain. In the strain-free state, our calculated friction coefficient is close to the value from Falk et al., 1.48 vs. 1.20 ($10^4$Ns/m$^3$) \cite{26}. The difference can be attributed to the different water models and the different LJ potential parameters used to describe the carbon-oxygen interactions (possibly yielding different contact angles). Compared to our previous MD simulations of water transport inside a double-walled CNT with a diameter of 2 nm \cite{23}, the friction coefficient of the graphene channel is higher, i.e., 1.48 vs. 0.3 ($10^4$Ns/m$^3$), which is consistent with the conclusion of Falk et al \cite{26} that $\lambda$ depends on the degree of curvature.
 
Since the slip length is a widely used quantity to describe the slip flow, we converted our friction coefficients $\lambda$ to the slip lengths $l_s$ by using $l_s = \mu/\lambda$ \cite{25}, where $\mu$ is the viscosity of water (0.82 mPaás for SPC/E water model at 300K \cite{27}). The results are plot in Fig. \ref{fig2}(b). At the strain-free state (i.e., $\epsilon = 0$), the slip length $l_s$ is about 54 nm. Such a large slip length in comparison to the channel height of 2 nm implies the plug-like speed profile shown in Fig. \ref{fig1}(b). We note that although our calculated slip length is significantly higher than the value obtained by Thomas et al. $\sim 30$ nm \cite{28}, the qualitatively form of the agreement is reasonable. We believe that the discrepancy arises from the different water model (TIP5P) adopted. Overall, however, the remarkable effect shown in these results is the degree to which strain engineering of the graphene planes can significantly change the slip length from 26 nm up to 173 nm.
 
The sensitive dependence of the friction coefficient $\lambda$ and the slip length $l_s$ on the strain (Fig. \ref{fig2}) suggests that the strain engineering indeed can serve as an effective method of controlling the water transport inside the graphene nano-channels. It may also be an important factor in our better understanding why there is such scatter in experimental and numerical results for slip lengths of water transport inside the CNTs; this can range from several nanometers up to several microns \cite{9, 10, 11, 29, 30}. To gain a deep physical insight into how the molecular mechanisms affect the drastic changes arising from the imposed strain, we investigated the phenomenon further, the results of that research is provided in the following sections.
 
\subsection{Molecular Structures of Water inside a Graphene Nano-channel}
It is known that the interfacial molecular structures determine the slip flow over a hydrophobic surface. In this section, we study the structural details of our water/graphene-channel system.
 
\begin{figure}[htbp]
\centerline{\includegraphics[width=0.8\textwidth]{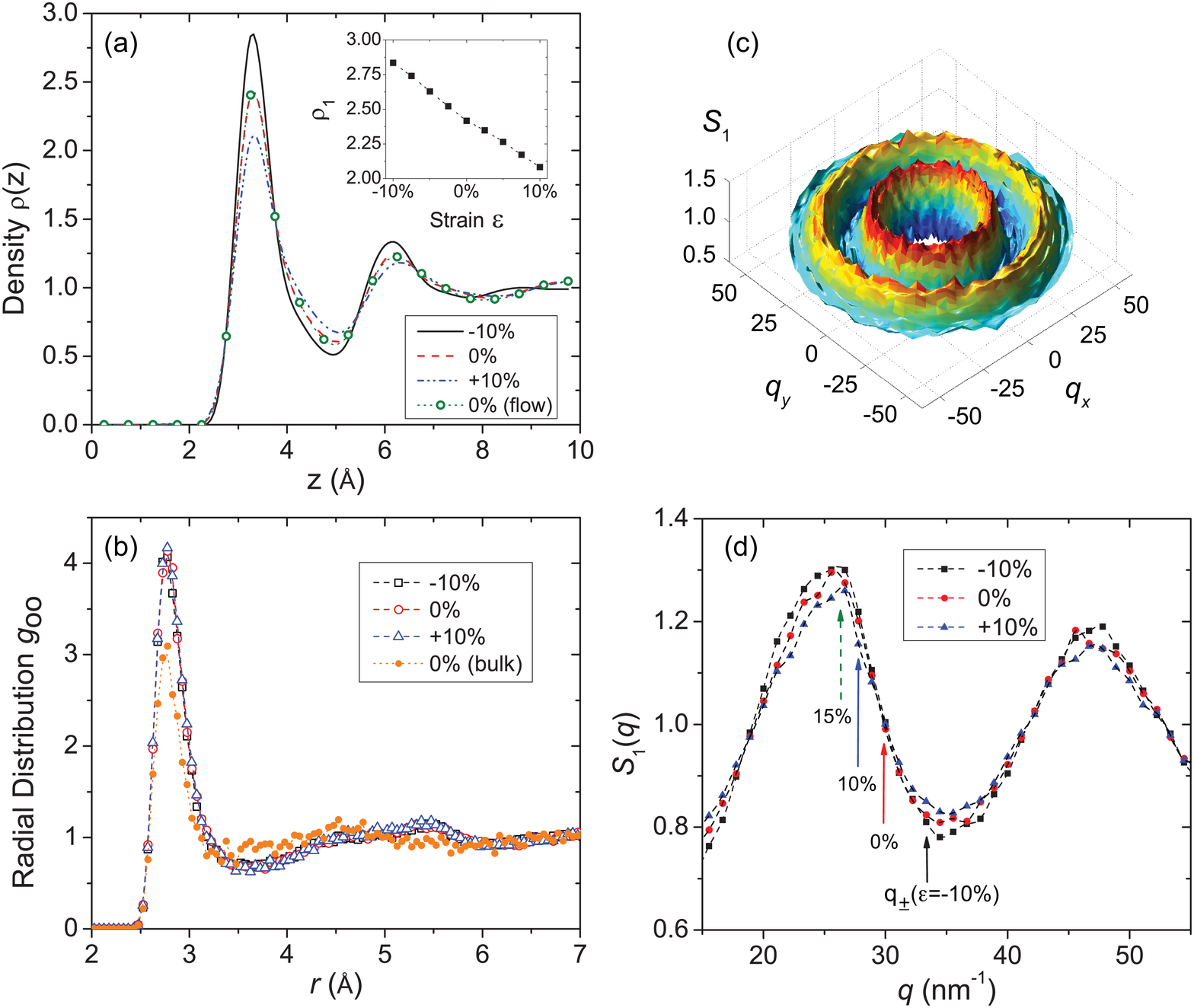}}
\caption{(color online) (a) Density profiles of water across the height ($z$ direction) of the graphene nano-channel. The sharp density peak next to the solid wall represents the 
first water layer. For graphene layers at zero strain, the density profiles at flow velocity $v_s=0$ m/s and $v_s=100$m/s have a negligible 
difference. The density profiles in the stained graphene channels with $\epsilon = -10\%$ (black), $0\%$ (red), $10\%$ (blue) indicate the almost same first peak position but different magnitude. Inset shows the density of the first peak as a function of the imposed strains $\epsilon$ on graphene. 
(b) Two-dimension radial density function (RDF, normalized by the 
mean density) of the first water layer in three differently strained graphene channels $\epsilon = -10\%$ (black), $0\%$ (red), $10\%$ (blue). In comparison, the RDF of the bulk water is also shown. (c) Two-dimensional structure factor $S_1(\textbf{q})$ of the first water layer in the strain free graphene channel exhibits a clear direction independence. (d) The radial average of the structure factor in the strained graphene channels with $\epsilon= -10\%$ (black), $0\%$ (red), and $10\%$ (blue). The solid arrows indicate the positions of the reciprocal lattice vector $|q_{\pm}|$ of the strained graphene planes.}
\label{fig3}
\end{figure}

Density profiles of the water molecules across the height of the graphene channels are shown in Figure \ref{fig3}(a). In the strain-free graphene channel, our MD results found  almost no difference between the densities of the water molecules at rest and at a flow rate of 100 m/s. The sharp peak indicates the position of the first liquid layer, i.e., $3.25 \text{\AA}$ from the solid walls. When different strains ($-10\%$ and $10\%$) are imposed on the graphene planes there is no change in the locations of the first density peaks, whereas the peak heights do decrease with the strain. The inset of Fig. \ref{fig3}(a) summarizes the density change with respect to the strains. We believe that the reason for the density drop is the weakened attraction between the solid walls and the water molecules, arising from the low surface density of the carbon atoms caused by the stretching the graphene layer.

The two-dimension radial density function (RDF) of the first liquid layer was also calculated. By normalizing the RDF with the mean density, we found that the RDFs almost overlap with each other for the differently strained graphene channels, as indicated in Fig. \ref{fig3}(b) at strains of $-10\%$, $0\%$, and $10\%$. This implies that strain engineering on the solid walls has a negligible effect on the structures of the first water layer. In comparison, the RDF of the bulk water is shown as the dashed line in Fig. \ref{fig3}(b). The fact that the first peak is in almost the position in this case suggests that the average oxygen-oxygen inter-atomic distance of the neighbored water molecules are unaffected in the graphene nano-channels, i.e., $r_{\text{OO}} = 0.275$ nm. We can thus conclude that the size confinement actually leads to more compact water molecules in the first liquid layer than in the bulk water without changing the inter-molecular distance. 

The 2-dimensional structure factor of the first water layer was calculated by using the following expression \cite{26, 31}
\begin{equation}
\label{eq1}
   S_1(\textbf{q}) = \langle \frac{1}{N}\sum_{j=1}^{N}\sum_{l=1}^{N}e^{i\textbf{q}\cdot(\textbf{x}_l-\textbf{x}_j)}\rangle \\
                           = \langle \frac{1}{N} \big[ \left( \sum_{j=1}^{N} \cos(\textbf{q}\cdot \textbf{x}_{j}) \right)^2+\left( \sum_{j=1}^{N} \sin(\textbf{q} \cdot \textbf{x}_{j}) \right)^2 ] \rangle                           
\end{equation}
where $\textbf{x}_j$ represents the position of the $j$th oxygen atom, $N$ is the number of oxygen atoms in the first water layer, and $\textbf{q}$ is the 2-dimensional reciprocal lattice vector. The bracket $\langle \rangle$ denotes an equilibrium ensemble average. Figure \ref{fig3}(c) shows the structure factor of water inside the zero-strained graphene channel. It is clear that the structure factor is almost independent of the direction of the $\textbf{q}$ vector. Similar isotropic structure factors are also observed in the strained graphene channels. We then averaged the $S_1(\textbf{q})$ along all the $\textbf{q}$ directions and plot the $S_1(q)$ in Fig. \ref{fig3}(d) with $q = |\textbf{q}|$ representing the length of the reciprocal lattice vector. The very small differences in the plots of $S_1(q)$ confirm our conclusion that the strain engineering on graphene walls does not affect the structures of the first water layer. However, the strain engineering does change the interfacial commensurateness due to the change of the graphene. The solid arrows in Fig. \ref{fig3}(d) represent the positions of the reciprocal lattice vectors of the differently strained graphene planes $|\textbf{q}_{\pm}|$. Here the lattice vectors of graphene are connecting the centers of two six-rings in neighbors. With the strains from $-10\%$ to $10\%$, it is evident that $S_1(|\textbf{q}_{\pm}|)$ increases, which suggests a better degree of commensurateness.

\begin{figure}[htbp]
\centerline{\includegraphics[width=0.8\textwidth]{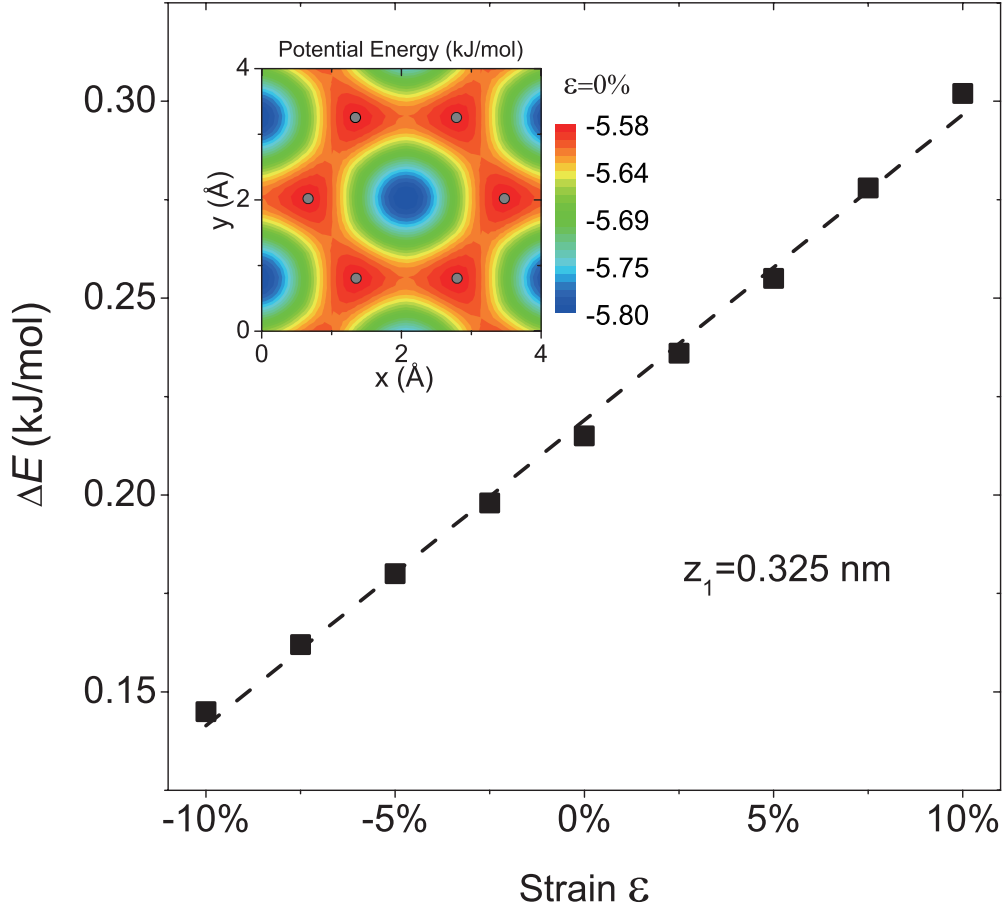}}
\caption{(color online) Interfacial potential energy barriers as a function of the strain applied on the graphene walls. The potential energy profile at $\epsilon = 0$ is shown in inset with a six-fold symmetry, which is consistent to the atomic structure of grapahene.}
\label{fig4}
\end{figure}

The inset of Fig. \ref{fig4} shows the interfacial potential energy profile over a graphene ($\epsilon = 0$) at the position of the first water layer, i.e., $z_1 = 3.25 \text{\AA}$ indicated in Fig. \ref{fig3}(a). The energy profile has a six-fold symmetry. In our analysis, the energy barrier $\Delta E(z_1)$ is defined as the difference of the potential energy between the maximum points (located on top of the carbon atoms) and the minimum points (located in the centre of the six-rings). Figure \ref{fig4} depicts a linear relation of the calculated energy barrier $\Delta E(z_1)$ with respect to the exerted strain. The magnitude of the energy barrier doubles when the strain $\epsilon$ changes from $-10\%$ to $10\%$.

\subsection{A Microscopic Understanding on Strain Dependent Friction}
As a dissipation coefficient, the friction coefficient $\lambda$ can be expressed via the Green-Kubo (GK) relationship, which relates $\lambda$ to the autocorrelation function of fluctuating pair-wise forces at equilibrium \cite{32}
\begin{equation}
\label{eq2} 
   \lambda = \frac{1}{Ak_BT}\int_{0}^{\infty}\langle \textbf{F}(t)\textbf{F}(0)\rangle dt 
\end{equation}
where $\textbf{F}(t)$ is the total forces in the flow direction exerted on carbon atoms due to the interactions with water molecules in our equilibrium MD simulations, $k_B$ is Boltzmann constant, $A$ is the surface area, and $T$ is water temperature. We directly calculated the force autocorrelation function $\langle \textbf{F}(t)\textbf{F}(0) \rangle$ in our equilibrium MD simulations with simulation time of 5 ns. Here, we should point out that a well-documented difficulty of estimating the GK relationships via the equilibrium MD is the finite size of the simulated system, which often leads to a vanishing of the friction coefficient after a very long time simulation \cite{32, 33}. The integration should thus have a cutoff time $t_0$. A widely adopted method of resolving this is to use the onset of a plateau of the integrations as the cutoff $t_0$. In our simulations, however, it was difficult to locate the plateau in some cases. So we followed the suggestions from Ref. \cite{33, 34} and chosen $t_0$  as the first zero of the force autocorrelation function, typically this was in the range [1ps, 10ps]. The friction coefficients calculated from our equilibrium MD simulations (Eq. (\ref{eq2})) are shown as the solid circles in Figure \ref{fig2}(b). A good agreement can be observed in comparison with the NEMD results, which suggests the GK relation is applicable in our systems. 

\begin{figure}[htbp]
\centerline{\includegraphics[width=0.8\textwidth]{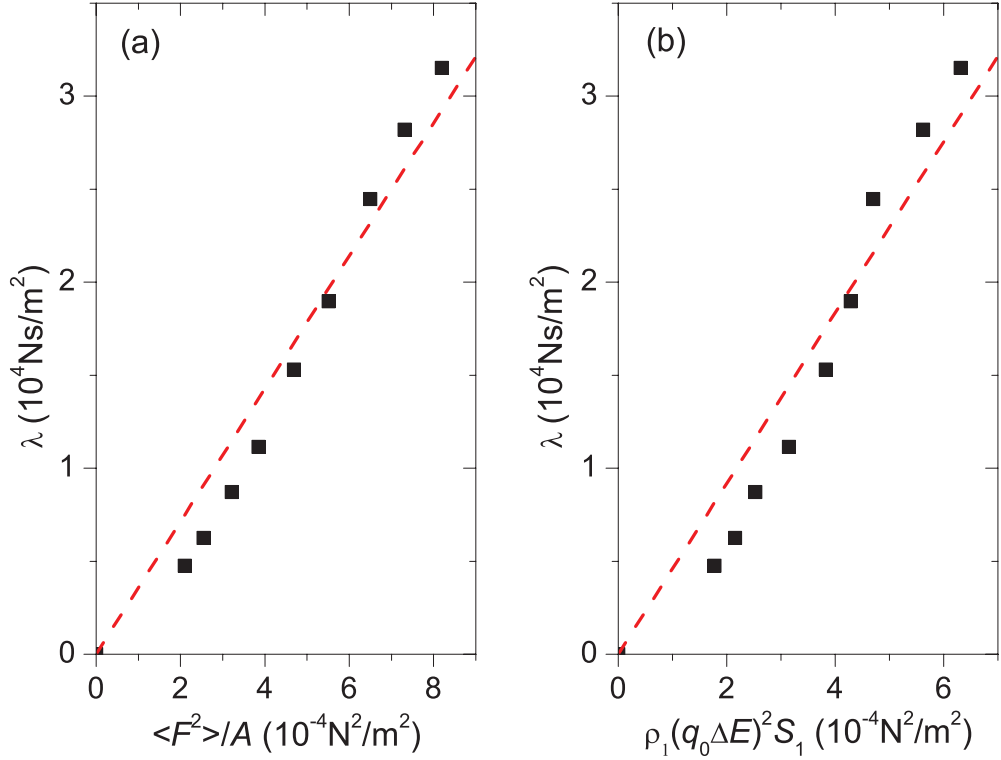}}
\caption{(a) Functional dependence of the friction coefficient $\lambda$ versus the static root mean square force $\langle F^2 \rangle$ (Eq. (\ref{eq3})). (b) Functional dependence of the friction coefficient $\lambda$ on the computed  $\rho_1(q_0\Delta E)^2 S_1$ (Eq. (\ref{eq4})). Dashed straight lines are guides to the eye. The good linear relations validate the forms of Eqns. (\ref{eq3}) and (\ref{eq4}).}
\label{fig5}
\end{figure}

To gain further insight of the molecular mechanisms, we consider the GK expression (Eq. (\ref{eq2})) in more details. The GK relation can be re-expressed as \cite{26}
\begin{equation}
\label{eq3} 
   \lambda = \frac{\tau_F}{Ak_BT}\langle \textbf{F}^2 \rangle.
\end{equation}
In our equilibrium MD simulations, we find that the de-correlation time $\tau_{F} = \int \langle \textbf{F}(t)\textbf{F}(0) \rangle / \langle \textbf{F}^2\rangle $, weakly depends on the exerted strains on the graphenes. Typically it is 100 -160 fs for $-10\% \le \epsilon \le 10\%$, which is consistent with the MD results of water flow in CNTs \cite{26}. Since the variance of $\tau_F$ (60$\%$) is one order of magnitude smaller than that of $\lambda$ (600$\%$), the main contribution to the variance of $\lambda$ must be from the static root mean square (RMS) force $\langle \textbf{F}^2 \rangle$. In other words, the change of the friction coefficients $\lambda$ with respect to the strains should be directly correlated with $\langle \textbf{F}^2 \rangle$. Indeed, as shown in Fig. \ref{fig5}(a), the good linear relation between $\lambda$ and $\langle \textbf{F}^2 \rangle /A$ suggests that the de-correlation time $\tau_F$ can be approximated as a constant.

Following Ref. \cite{26}, the RMS force $\langle \textbf{F}^2 \rangle$ can be estimated analytically. If we assume the main contribution of the friction shear stress arises from the first liquid layer, one can approximate the total RMS force as:
\begin{equation}
\label{eq4} 
   \frac{\langle \textbf{F}^2 \rangle}{A} \cong \frac{1}{2}\rho_1 \left(S_1(\textbf{q}_{+})+S_1(\textbf{q}_{-}) \right) \left(q_0 \Delta E \right)^2
\end{equation}
where $\rho_1$ is the density of the first water layer, $S_1$ is the 2-dimensinonal structure factor of the first water layer, and $\Delta E= \Delta E(z_1)$ is the energy barrier. The reciprocal lattice vector of the graphene wall is $\textbf{q}_{\pm}=q_0(1/\sqrt{3}; \pm1)$ with $q_0 = 2\pi/(\sqrt{3}a_{\text{CC}})$ and $a_{\text{CC}}$ the carbon-carbon bond length. Figure \ref{fig5}(b) shows the comparison of the friction coefficients directly calculated from our MD simulations (Fig. \ref{fig2}(b)) and those computed by using the analytical expressions in Eq. (\ref{eq4}). The good linear relation clearly validates the form of Eq. (\ref{eq4}). This is an important result because we can quantitatively analyze the effects of density $\rho_1$, structure factor $S_1(\textbf{q}_{\pm})$, and energy barrier $\Delta E$ on the friction coefficients.

\begin{figure}[htbp]
\centerline{\includegraphics[width=0.8\textwidth]{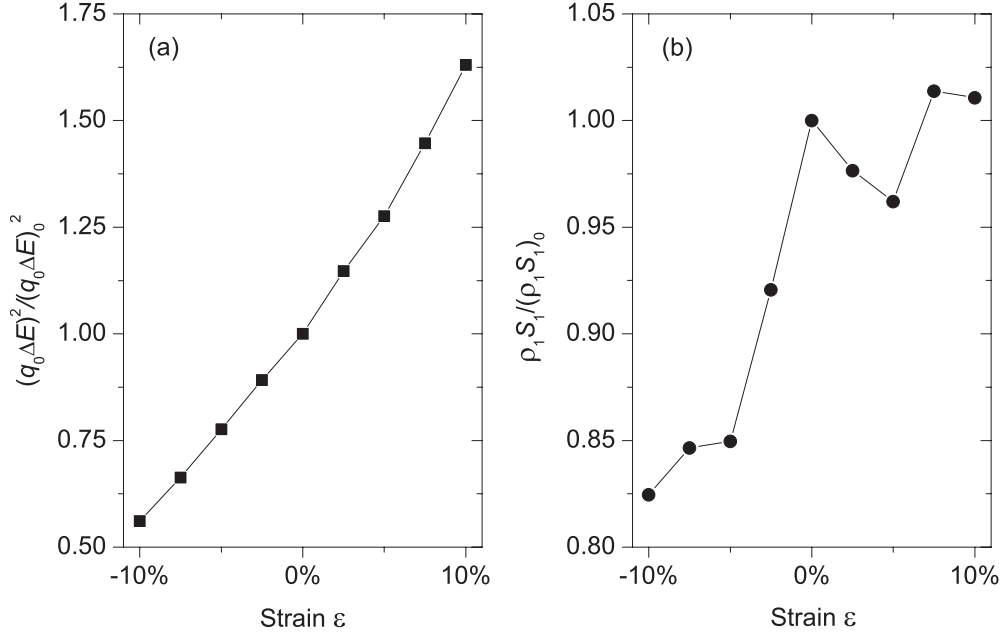}}
\caption{(a) The normalized energy barrier of the potential profiles from the graphene solid walls as a function of the strains applied. (b) The normalized structural changes as the function of the applied strains on the solid walls.}
\label{fig6}
\end{figure}

To quantify the contributions of the energy barrier and the interfacial structures to the friction coefficient $\lambda$, we normalize $\lambda$ by the friction coefficient at zero strain $\lambda_0$ as $\lambda/\lambda_0 = [(q_0\Delta E)^2/(q_0\Delta E)_{0}^{2}][\rho_1S_1/(\rho_1S_1)_0]$. Figure \ref{fig6} shows how each of these two terms varies as functions of the strain $\epsilon$ applied on graphene walls. When compared to Fig. \ref{fig2}(b), we can conclude that: first, the contribution to the change of $\lambda$ mainly comes from the change of the energy barrier (Fig. \ref{fig6}(a)); second, in the stretched graphene nano-channels, the increase of commensurateness $S_1$ (Fig. \ref{fig3}(d)) and the reduction of the first liquid layer density $\rho_1$ (Fig. \ref{fig3}(a) inset) cancel out each other, resulting in a small overall contribution (Fig. \ref{fig6}(b)); third, when the graphene channel under in-plane compression, the decrease of the structural factor overwhelms the increase of the density (Fig. \ref{fig6}(b)), meaning that, quantitatively, the contribution from $\rho_1 S_1$ is about $30\%$ of that from $\Delta E$ towards the effect on the friction coefficient (Eq. (\ref{eq4})).

\section{Conclusions}
To summarize, our MD simulations show that the friction coefficient $\lambda$ (and the slip length $l_s$) of the water transport in the graphenes nanochannels exhibits a highly sensitive dependence on the strains imposed on the graphene. The friction coefficients change by a factor of 6 when the strain on the graphene wall changes from $-10\%$ to $10\%$. This corresponds to a change in the slip length $l_s$ varies from 26 nm to 173 nm. Our results suggest that strain engineering could serve as an effective route to control the water transport inside graphene nanochannels. It may also be an important factor in understanding the scatter in the reported slip lengths of the water flow inside CNTs in experiments and simulations. The molecular mechanisms of the strain effect on the slip flow are also studied. We find that the  strains on the graphene have relatively small influences on the molecular structure of the first water layer, other than on the reduction of the density. Using a simplified analytical model based on the Green-Kubo relation [26], we find that the change of energy barrier makes the most important contribution to the change of the friction coefficient $\lambda$, in comparison to the effect of the water density and the structural factor (which make about $30\%$ of the contribution made by the energy barrier). The quantitative relationship for $\tau - v_s$ provided herein when combined with the physical insights provided on the molecular mechanism will be valuable to designers of graphene nano-channels for application in nano-fluidic devices.

%In the positive strain range, the contributions from the interfacial commensurateness, i.e., structure factor $S_1$ and the density $\rho_1$ of the first water layer almost cancel out each other. In the negative strain range, the change of structural factor overwhelms the density change and thus the overall contribution is about $30\%$ of that of the energy barrier to the flow friction coefficient $\lambda$. The obtained quantitative $\tau - v_s$ relations and the revealed molecular mechanism will be valuable to design the graphene nano-channels in the nano-fluidic devices.

%In addition, recalling that the IHS relationship is derived based on transition state theory and velocity $\tau/\tau_0 = asinh(v_s/v_0)$, it indicates that interfacial water flow in nano-scale is a diffusive process instead of a viscous flow. The physical meanings of fitting parameters $\tau_0$ and $v_0$ (Table I) are not well understood. Our study, i.e., strain engineering on the graphene solid walls, could serve as a good system to reveal the physical meanings of the two quantities. This could be an interesting research in the future.

Q.S.Z. acknowledges the financial support from NSFC through Grant No.
10832005, the National Basic Research Program of China (Grant No.
2007CB936803), and the National 863 Project (Grant No. 2008AA03Z302). J.Z.L.
acknowledges new staff grant 2010 and small grant 2011 from engineering
faculty of Monash University. This work was supported by an award under the Merit Allocation Scheme on the Australia NCI National Facility at the ANU.

% ------------------------------------------------------------------------
%\nocite{*}
% ------------------------------------------------------------------------
%Included for Gather Purpose only:
%\bibliographystyle{unsrt}
\pagebreak

\pagebreak

\end{document}